\documentclass[conference]{IEEEtran}
\usepackage{graphicx}
\usepackage{subfigure}
\usepackage{float}
\usepackage{amsmath}
\usepackage{amsfonts,amssymb}
\usepackage{dsfont}
\usepackage{algpseudocode}
\usepackage{makecell}
\usepackage{cite}
\usepackage{multirow}
\usepackage{multicol}
\usepackage{booktabs}
\usepackage[ruled,vlined, linesnumbered,lined,boxed,commentsnumbered,ruled,longend,noend]{algorithm2e}
\usepackage{xcolor,soul}
\usepackage{comment}
\usepackage{bbm}

\newcommand\myeq{\mathrel{\stackrel{\makebox[0pt]{\mbox{\normalfont\tiny (3), (4)}}}{=}}}

\SetKwInput{KwInput}{Input}                
\SetKwInput{KwOutput}{Output}              

\usepackage{fancyhdr,lipsum}

\fancypagestyle{mahmood}{%
   \fancyhf{} 
   
   \fancyhead[C]{You can use this material personally. Reprinting or republishing this material for the purpose of advertising or promotion, creating new collective works, reselling or redistributing to servers or lists, or using any copyrighted component in other works must adhere to IEEE policy. The DOI will be supplied as soon as it becomes available.}
}%

\makeatletter
\let\ps@IEEEtitlepagestyle\ps@mahmood
\makeatother

\begin{document}
\title{Semantic-Aware Dynamic and Distributed Power Allocation: a Multi-UAV Area Coverage Use Case}

\author{
    \IEEEauthorblockN{
        Hamidreza Mazandarani\textsuperscript{1}, Masoud Shokrnezhad\textsuperscript{2}, and Tarik Taleb\textsuperscript{1} \\
    }
    \IEEEauthorblockA{
       \textsuperscript{1} \textit{Ruhr University Bochum, Bochum, Germany; hr.mazandarani@ieee.org, tarik.taleb@rub.de} \\
        \textsuperscript{2} \textit{ICTFicial Oy, Espoo, Finland; masoud.shokrnezhad@ictficial.com}
    }
}

\maketitle

\begin{abstract}
The advancement towards 6G technology leverages improvements in aerial-terrestrial networking, where one of the critical challenges is the efficient allocation of transmit power. Although existing studies have shown commendable performance in addressing this challenge, a revolutionary breakthrough is anticipated to meet the demands and dynamism of 6G. Potential solutions include: 1) semantic communication and orchestration, which transitions the focus from mere transmission of bits to the communication of intended meanings of data and their integration into the network orchestration process; and 2) distributed machine learning techniques to develop adaptable and scalable solutions. In this context, this paper introduces a power allocation framework specifically designed for semantic-aware networks. The framework addresses a scenario involving multiple Unmanned Aerial Vehicles (UAVs) that collaboratively transmit observations over a multi-channel uplink medium to a central server, aiming to maximise observation quality. To tackle this problem, we present the Semantic-Aware Multi-Agent Double and Dueling Deep Q-Learning (SAMA-D3QL) algorithm, which utilizes the data quality of observing areas as reward feedback during the training phase, thereby constituting a semantic-aware learning mechanism. Simulation results substantiate the efficacy and scalability of our approach, demonstrating its superior performance compared to traditional bit-oriented learning and heuristic algorithms.
\end{abstract}


\section{Introduction}
A myriad of technological advancements synergistically contributes to the global transition to sixth-generation (6G) mobile networks. Among these advancements, aerial-terrestrial networking holds the promise of providing more widespread access, particularly to previously unconnected regions of the world \cite{drones6060147, gao2024space}. This networking approach integrates aerial platforms, such as Unmanned Aerial Vehicles (UAVs), with traditional terrestrial communication infrastructures, thereby offering flexible and on-demand connectivity in areas where deploying fixed infrastructure is either challenging or economically unfeasible. For example, UAVs can be deployed to facilitate remote monitoring tasks. Notably, applications such as wildlife counting \cite{attard2024review} and intelligent transportation systems \cite{afrin2024advancements} stand to benefit significantly from this integration. To optimize performance, especially in diverse and dynamic environments in which aerial-terrestrial networks operate, a critical challenge is the efficient allocation of user transmit power. This involves meticulously adjusting the transmit power level of each user to ensure that a predefined Quality of Service (QoS) threshold is achieved at the receiver. The power allocation process requires a comprehensive analysis of the communication environment, taking into account factors such as noise and, particularly, interference from other users sharing the same channel.

Several noteworthy approaches have emerged among studies that tackle the transmit power allocation problem in aerial-terrestrial networks. For instance, Fu \textit{et. al.} \cite{fu2023uav} focused on power control in energy-harvesting UAVs, integrating trajectory management and user association. They proposed an offline approach based on successive convex approximation, alongside an online convex-assisted Reinforcement Learning (RL) method. Alnakhli \textit{et. al.} \cite{10520245} explored bandwidth allocation and power control within a multi-UAV network, employing decomposition and sequential quadratic programming techniques. Li \textit{et. al.} \cite{li2024blocklength} examined the joint allocation of blocklength and power control in uplink UAV-assisted networks and proposed a multi-agent RL process. Yuan \textit{et. al.} \cite{10496864} investigated the joint problem of power control and UAV trajectory design, employing dual-primal theory along with mechanical rope equilibrium. Lastly, Ning \textit{et. al.} \cite{10537097} intended to optimize joint user association, interference cancellation, and power control for UAV communications, utilizing inverse soft-Q learning and successive convex approximation techniques. Although these approaches demonstrate exceptional results, they are often overly complex and may not be efficient enough to meet the substantial demands of 6G networks that require scalable performance. Furthermore, most of these methods rely on offline optimization-based techniques, rendering them rigid in the highly dynamic scenarios characteristic of 6G environments.

To address the issue of performance scalability, a potential candidate is the paradigms of semantic communication and orchestration, which enhance network awareness of content meaning and purpose \cite{shokrnezhad2024semantic}. Semantic communication shifts the focus from merely transmitting raw bits to conveying the intended meanings and interpretations of data. Semantic orchestration involves the intelligent coordination of network resources to ensure that semantic information is effectively utilized across the entire network. These paradigms dramatically improve resource utilization efficiency by compressing transmitted data and facilitating semantic-aware resource sharing \cite{mazandarani2024semantic}. To address the adaptability issue (while maintaining scalability), state-of-the-art distributed Machine Learning (ML) techniques can be employed, which enable centralized training with decentralized execution. This approach allows agents to learn from a collective experience while operating independently in their respective environments. These distributed methods are particularly pertinent in complex scenarios, such as the Metaverse, where rapid changes and diverse data sources necessitate a balance between scalability and the efficiency of learning solutions \cite{taleb2023ai, 10816182}.

In light of this context, this paper focuses on investigating the problem of semantic-aware distributed power allocation in multi-channel wireless networks. The use case being examined involves multiple UAVs engaged in area coverage, whereby a group of UAVs navigates over a designated region and collaboratively transmits their observed scenes to a central server connected to all ground BSs. To address this problem, we implement a customized version of the Semantic-Aware Multi-Agent Double and Dueling Deep Q-Learning (SAMA-D3QL) algorithm, which was developed in our previous research efforts \cite{mazandarani2024semantic}. This approach leverages a Deep Reinforcement Learning (DRL) framework in which the reward structure is designed to reflect the data quality of all observation areas, while the state space incorporates metrics such as the shared coverage degree. This effectively constitutes a semantic-aware learning mechanism, promoting optimal power allocation and enhancing overall system performance. Using SAMA-D3QL, UAVs dynamically and distributedly adjust their transmission power to construct the highest possible quality image on the server, learning to manage observation overlaps effectively to minimize unnecessary interference.

The remainder of this paper is organized as follows. Section \ref{s_prb_stt} presents the system model and problem formulation. Section \ref{s_prp_slt} elucidates the proposed solution, encompassing SAMA-D3QL elements and learning architecture. In Section \ref{s_sim}, we present and analyze numerical results for several environmental attributes. Finally, Section \ref{s_con} concludes the paper with a summary of our findings and closing remarks on the implications and potential future research directions.

\vspace{-4pt}
\section{Problem Statement}\label{s_prb_stt}

\subsection{System Model}\label{ss_sys_mdl}
\begin{figure}[!t]
    \centerline{\includegraphics[width=3.6in]{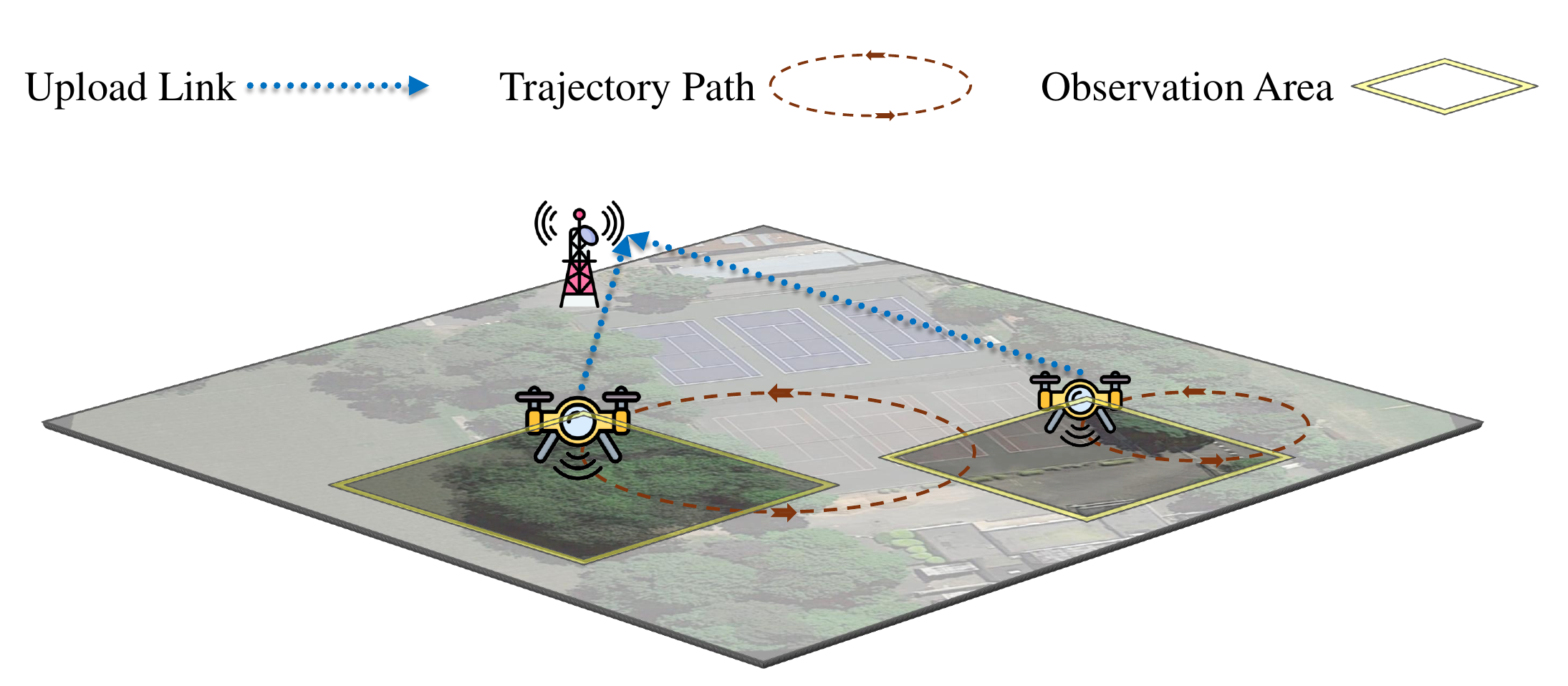}}
    \caption{A simple scenario featuring a single BS and two UAVs hovering in circular paths, transmitting their observations to the BS.}
    \vspace{-10pt}
    \label{fig1}
\end{figure}

This paper focuses on an aerial wireless network characterized by $\mathcal{B}$ ground BSs and $\mathcal{N}$ UAVs, denoted as $\mathbb{B}$ and $\mathbb{N}$, respectively. Fig. \ref{fig1} illustrates a sample scenario with $\mathcal{B} = 1$ and $\mathcal{N} = 2$. We exclusively consider uplink transmissions from UAVs to BSs, and the data rate for each UAV $i$ over a set $\mathbb{C}$ of $\mathcal{C}$ channels is calculated using the Shannon-Hartley theorem, as follows:
\begin{align} \label{rates}
    \dot{r}_{i}^{t} &= \sum_{c \in \mathbb{C}}^{}\log_{2}{ \Big( 1 + \gamma_{i, c}^{t} \Big) }\notag \\
    &= \sum_{c \in \mathbb{C}}^{}\log_{2}{ \Big( 1 + \frac{p_{i, c}^{t} \cdot h_{i, {b}_{i}^{t}}^{t}}{\sum\limits_{j \in \mathbb{N} \setminus \{i \} }^{}{p_{j, c}^{t} \cdot h_{j, {b}_{i}^{t}}^{t}} + {\sigma}_{c}^{2}} \Big) }.
\end{align}
In this formulation, the following variables and parameters are defined:
\begin{itemize}
    \item \(\gamma_{i, c}^{t}\) denotes the Signal-to-Interference-plus-Noise Ratio (SINR) of UAV \(i\) at BS \(b_{i}^{t}\) during time slot \(t\) on channel \(c\). SINR is a critical metric that quantifies the quality of a wireless communication link, measuring the ratio of the desired signal power to the sum of interference power from other transmitting sources and the background noise power. 
    \item $p_{i, c}^{t}$ denotes the transmit power of UAV $i$ at time slot $t$ on channel $c$, and $\mathbb{P} \in \mathbb{R}^{\mathcal{N} \times \mathcal{C} \times \mathcal{T}}$ represents the vector encompassing the transmit power of all UAVs at all channels and all time slots.
    \item $h_{i, {b}_{i}^{t}}^{t}$ signifies the channel-independent path gain between UAV $i$ and BS ${b}_{i}^{t}$ at time slot $t$. 
    We consider Free Space path gain between UAV $i$ located at $l_{i}^{t} = (x_{i}^{t}, \ y_{i}^{t}, \ z)$ and BS $b$ located at $\bar{l}_{b}^{t} = (\bar{x}_{b}^{t}, \ \bar{y}_{b}^{t}, \ \bar{z})$ during time slot $t$, therefore:
    \begin{align}\label{eq_prb}
       h_{i, b}^{t} = \Big( \sqrt[]{(x_{i}^{t} - \bar{x}_{b}^{t})^2 + (y_{i}^{t} - \bar{y}_{b}^{t})^2 + (z - \bar{z})^2} \Big)^{-\alpha}.
    \end{align}
    For the purposes of this study, it is assumed that each UAV is connected to the BS with the highest path gain.
    \item ${\sigma}_{c}^{2}$ represents the noise power on channel $c$.
\end{itemize}

Assuming that UAVs' trajectories are predetermined and treated as input for our system model, we consider that each UAV observes a square area during each time slot \(t\). The side length of this square area observed by UAV \(i\) is denoted as \(d_{i}\). Therefore, the data quality of each UAV is calculated as follows, where $\Lambda(.)$ is the mapping of UAV rates per square meter to image qualities, as follows:
\begin{align}\label{quality}
    q_{i}^{t} &= \Lambda\Big( \frac{\dot{r}_{i}^{t}}{{d}_{i}^{2}} \Big).
\end{align}
\(\Lambda(.)\) is not known in advance; therefore, it needs to be learned by UAVs. We assume that each range of supported data rates corresponds to a specific bit rate for image quantization. In other words, higher data transmission rates allow for higher bit rates in image quantization, which in turn results in better-quality images. To evaluate the quality of the transmitted images, we compare the Peak Signal-to-Noise Ratio (PSNR) metric between the quantized image and the original image. The PSNR serves as a measure of image quality (i.e., function $\Lambda(.)$), where higher PSNR values indicate a closer resemblance between the quantized image and the original, thus reflecting better transmission quality.

\begin{figure}[!t]
    \centerline{\includegraphics[width=1.8in]{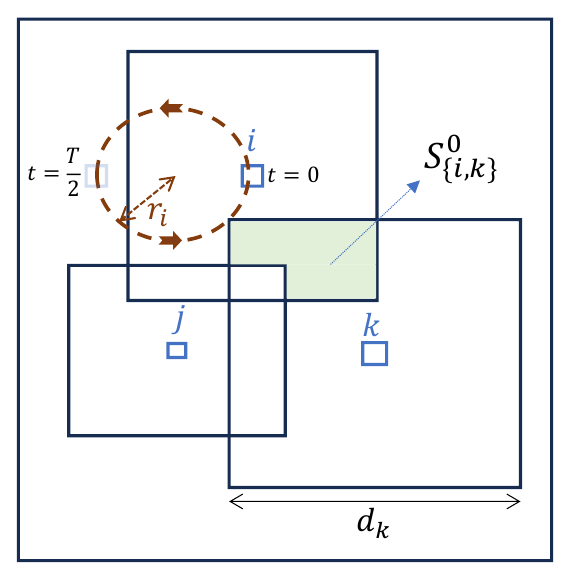}}
    \caption{At time \(t = 0\), three UAVs—denoted as \(i\), \(j\), and \(k\)—create a total of seven observing segments, such as \(S_{{i, k}}^{0}\). The dashed red line represents the circular trajectory of UAV \(i\), illustrating its path while observing the area.}
    \vspace{-10pt}
    \label{segment_modeling}
\end{figure}

Since the coverage areas of UAVs overlap, we define a set of segments \(\mathbb{S} = \{ S \subseteq \mathbb{N} \}\), where each segment \(S\) is associated with a specific group of UAVs that cover that area. An example of this concept is illustrated in Fig. \ref{segment_modeling}. The data quality of each segment is determined by the UAV within that segment that produces the highest quality image. This is formally expressed in \eqref{segment_quality}. The rationale behind this approach is that the BSs prioritize images captured by UAVs with the highest data transmission rates, as these are more likely to correspond to superior image quality.
\begin{align}\label{segment_quality}
    {Q}_{S}^{t} &= \max_{i}\Big\{ q_{i}^{t} | i \in S \Big\}
\end{align}

\subsection{Problem Formulation}\label{ss_prb_frm}
Defining the performance of each individual UAV in \eqref{quality}, the overall system performance can be characterized by the data quality of all the observing areas over time, which we denote as \(\mathcal{O}^{T}\), that is
\begin{align}\label{eq_obj}
\mathcal{O}^{T} &= \sum_{t \in \mathbb{T} }^{}{\sum_{S \in \mathbb{S} }^{}}{\chi_{S}^{t} \cdot {Q}_{S}^{t}} \notag \\
&\myeq \ \sum_{t \in \mathbb{T} }^{}{\sum_{S \in \mathbb{S} }^{}}{\chi_{S}^{t} \cdot \max_{i} \big\{ \Lambda( \frac{\dot{r}_{i}^{t}}{{d}_{i}^{2}} ) | i \in S \big\} },
\end{align}
where $\mathbb{T} = \{1, \ldots, \mathcal{T} \}$, and $\chi_{S}^{t}$ is the size of segment $S$ (per square meter). The sizes of the segments are calculated using straightforward numerical methods, such as basic geometric calculations, that take into account the corners of the coverage areas of UAVs.

Now, we can formulate the problem SPACE, which stands for \textbf{S}emantic-aware \textbf{P}ower \textbf{A}llocation for area \textbf{C}overag\textbf{E}. The goal of this problem is to optimize the overall system performance by adjusting the transmit power of UAVs. In doing so, we must take into account the constraints related to their data transmission rates as well as the limits on their transmit power, defined by \(\mathcal{P}_{\text{min}}\) (minimum power) and \(\mathcal{P}_{\text{max}}\) (maximum power).
\begin{align}\label{eq_prb}
    \text{SPACE:} \quad & \max_{\mathbb{P}}{\mathcal{O}^{T}} \quad \textit{s.t. Rate Constraints} \ \eqref{rates} \notag \\
     \textit{where: \ } &\mathcal{P}_{\text{min}} \leq \sum_{c \in \mathbb{C}}^{}{p_{i, c}^{t}} \leq \mathcal{P}_{\text{max}} \quad \forall i \in \mathbb{N}, \forall t \in \mathbb{T} \notag
\end{align}

\subsection{Complexity Analysis}\label{ss_cmp_ana}
SPACE is demonstrated to be NP-hard through a reduction from the well-known NP-hard Knapsack problem. In SPACE, the objective is to maximize the overall system performance of UAVs while satisfying specific rate constraints and power limitations. Each UAV's power allocation can be compared to selecting items in the Knapsack problem, where each item (represented by a UAV) has an associated value (the data quality achievable at a particular transmit power) and a weight (the power consumption of that UAV). The constraints in SPACE, specifically \(\mathcal{P}_{\text{min}} \leq \sum_{c \in \mathbb{C}} p_{i, c}^{t} \leq \mathcal{P}_{\text{max}}\), restrict the total transmit power utilized by UAVs, analogous to the weight limit imposed on selected items in the Knapsack problem. By finding an optimal power allocation strategy in SPACE that maximizes data quality, we fundamentally provide a solution to the Knapsack problem as well. Since the Knapsack problem is established as NP-hard, this reduction confirms that SPACE is likewise NP-hard, indicating that it is unlikely that a polynomial-time algorithm exists to solve it efficiently. This motivates us to adopt learning algorithms, particularly multi-agent learning algorithms, to address the dual aims of scalability and adaptability.

\section{Proposed Solution}\label{s_prp_slt}
This section explains the fundamental components of our Multi-Agent Deep Reinforcement Learning (MADRL)-based algorithm, built upon Semantic-Aware Multi-Agent Double and Dueling Deep Q-Learning (SAMA-D3QL) \cite{mazandarani2024semantic} and customized to solve SPACE in a distributed and dynamic manner. This exposition will cover discussions on action and state spaces, reward mechanisms, and the architecture of the learning process.

\subsubsection{Action Space}\label{ss_action_space}
The joint action space of all UAVs is specified in \eqref{action_space}. In this set, the action for UAV $i$ involves selecting transmitting power for each of the $\mathcal{C}$ channels from a discrete set of valid power levels denoted with $P_{Q}$, subject to the minimum and maximum sum constraints across all channels\footnote{For instance, with $P_{Q} = [0, 5, 10]$, $\mathcal{C} = 2$, $\mathcal{P}_{\text{min}} = 0$ and $\mathcal{P}_{\text{max}} = 10$, possible actions for each user is [[0, 0], [0, 5], [5, 0], [5, 5], [0, 10], [10, 0]].}.
\begin{equation}\label{action_space}
\boldsymbol{\alpha} = \bigg\{ \alpha_{i}: \big\{ p_{i, c} \in P_{Q} | \mathcal{P}_{\text{min}} \leq \sum_{c \in \mathbb{C}}^{}{p_{i, c}} \leq \mathcal{P}_{\text{max}}  \big\} \bigg| i \in \mathbb{N} \bigg\}
\end{equation}

\subsubsection{State Space}\label{ss_state_space}
Global system state at time slot $t$ entails histories (with length $\mathcal{H}$) of UAVs' local observations of their path gains and their locations in the grid, in addition to a parameter dubbed shared coverage degree (denoted with $\bar{z}_{i}^{t}$ for UAV $i$, defined as the average number of observing UAVs in $S_{i}^{t}$). Remarkably, all observation elements are normalized to their maxima.
\begin{align}\label{state_space}
& O_{i}^{t} = \{ h_{i, {b}}^{t} | b \in \mathbb{B} \} \cup \{ \bar{z}_{i}^{t} \} \cup \{ l_{i}^{t} \} \notag \\
& \boldsymbol{S}^{t} = { \bigg\{ {s}_{i}^{t} : \Big\{ O_{i}^{h} | h \in \{ t-\mathcal{H}, \ldots, t-1 \} \Big\} \ \bigg| \ i \in \mathbb{N} \bigg\} }
\end{align}

\subsubsection{Reward}\label{ss_reward}
As we consider a cooperative application in which all UAVs strive to maximize the data quality of all observing areas, the system reward is empirically defined as the objective function of SPACE, defined in \eqref{eq_obj}. Structuring the reward in this manner effectively constitutes a semantic-aware learning mechanism, as the reward system is intricately linked to the semantic understanding of the data, instead of merely optimizing for raw performance metrics, such as data rate.  

\subsubsection{Architecture}\label{ss_architecture}

SAMA-D3QL is grounded in the Value Decomposition Network (VDN) framework \cite{sunehag2017value} with CTDE strategy. This approach operates on the assumption that during the training phase, there is complete access to the global system state, while user decisions are made by their local information. In our approach, the training process involves a centralized server that connects to all BSs, where the policies for UAVs are being developed. However, it is assumed that BSs periodically update and distribute policies to UAVs via high-bandwidth and low-latency dedicated communication channels. UAVs autonomously select their actions on transmission powers based on their respective policies and individual observations. Using Q-learning, the summation of individual user Q-values is employed to compute the total Q-value within VDN, precisely calculated using the following equation:
\begin{align}\label{training}
Q_{tot}(\boldsymbol{S}^{t}, \boldsymbol{\alpha}^{t}) = \sum_{i \in \mathbb{N}}{}{Q_{i}({s}_{i}^{t}, \alpha_{i}^{t})}
\end{align}

With the above definition, the policy update function for all agents, represented as $\boldsymbol{\mathcal{W}}  = \{ \mathcal{W}_{i} \ | \ i \in \mathbb{N} \}$, can be expressed as follows:
\begin{align}
\label{eq_DQL_bellman}
\boldsymbol{\mathcal{W}}^{t+1} = & \boldsymbol{\mathcal{W}}^{t} \nonumber \ + \\
& \sigma[Y^t_{D3QL} - Q_{tot}(\boldsymbol{S}^{t}, \boldsymbol{\mathbb{\alpha}}^{t}; \boldsymbol{\mathcal{W}}^t)]\nabla_{\boldsymbol{\mathcal{W}}^{t}} \nonumber \ \cdot \\
& Q_{tot}(\boldsymbol{S}^{t}, \boldsymbol{\mathbb{\alpha}}^{t}; \boldsymbol{\mathcal{W}}^{t}),
\end{align}
where the target value ($Y^t_{D3QL}$) is determined using the D3QL algorithm, which is an extension of the original DQL algorithm that incorporates both dueling and double mechanisms \cite{mazandarani2024semantic}. Our approach is outlined in Algorithm \ref{alg_sama_d3QL}, wherein lines 5-11 pertain to the decentralized execution of actions by UAVs during training and test (or inference) phases, while lines 14-19 depict the centralized training conducted at the server during only the training phase. $\mathbb{E}_{train}$ and $\mathbb{E}_{test}$ refer to sets of training and test episodes, respectively.

\begin{algorithm}[t!]\label{alg_sama_d3QL}
\caption{\small{Semantic-aware Dynamic and Distributed Power Allocation}}
\KwInput{$\mathcal{T}, \mathbb{E}_{train}, \mathbb{E}_{test}$, UAV and BS locations ($L$)}
$\boldsymbol{\mathcal{W}} \leftarrow \mathbf{0}$, $\boldsymbol{\mathcal{W}}^{-} \leftarrow \mathbf{0}$, $\epsilon \gets 1$, $memory \gets \{\}$ \\
\ForEach{$ep$ in $\mathbb{E}_{train} \cup \mathbb{E}_{test}$}
{
    \ForEach{$t$ in $\mathbb{T}$}
    {
        \textcolor{gray}{$\star$ Decentralized Execution (5-11)}  \\
        \ForEach{$i$ in $\mathbb{N}$}
        {
            $\zeta \gets$ generate a random number from $[0:1]$ \\
            \If{$\zeta > \epsilon$}
            {
                $\alpha_{i}^{t} \gets$ argmax$_{\alpha \in \boldsymbol{\alpha}_{i}} Q({s}_{i}^{t}, \alpha, {\mathcal{W}}_{i})$ \\
            }
            \Else
            {
                select a random $\alpha_{i}^{t}$ from $\boldsymbol{\alpha}_{i}$
            }
            observe $\dot{r}_{i}^{t}$ and construct ${s}_{i}^{t+1}$ acc. to \eqref{state_space} \\
        }
        \If{$ep \in \mathbb{E}_{train}$}
        {
            \textcolor{gray}{$\star$ Centralized Training (14-19),} \\
            calculate $\boldsymbol{R}^{t}$ according to \eqref{eq_obj} \\
            $memory \gets \{ \boldsymbol{R}^{t} \} \cup \{({s}_{i}^{t}, \alpha_{i}^{t}, {s}_{i}^{t+1}) \big| \ i \in \mathbb{N} \}$ \\
            choose a batch of samples from $memory$\\
            train the agent according to \eqref{eq_DQL_bellman} \\
            \If{$\epsilon > \widetilde{\epsilon}$}
            {
                $\epsilon \gets \epsilon \cdot \epsilon'$
            }
        }
    }
}
\end{algorithm}

\section{Evaluation}\label{s_sim}
\subsection{Setup}
\begin{figure*}[!t]
\centerline{\includegraphics[width=\textwidth, height=3cm]{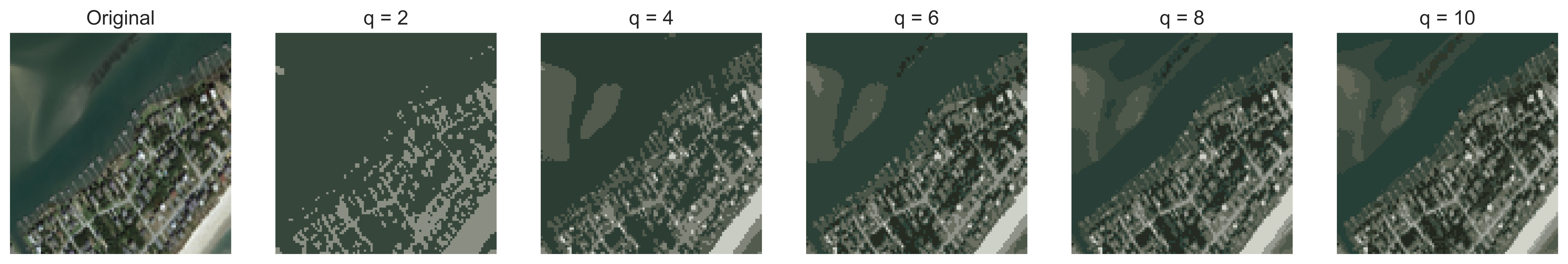}}
    \caption{Selected Image from the DOTA dataset \cite{xia2018dota}, and its quantized versions with different bits.}
    \label{plane_sample_quantized}
\end{figure*}

In this section, we present a numerical analysis of the proposed SAMA-D3QL-based solution using the parameters detailed in Table \ref{t_prm}. We investigate four scenarios: first, analyzing how the number of channels (as an indicator of interference boundedness) impacts performance; second, examining the effect of UAV count on performance to evaluate scalability; third, assessing system performance by varying UAV velocities to measure dynamicity; Finally, evaluating how different evaluation metrics influence performance. Evaluation planes are six aerial images selected from the DOTA dataset \cite{xia2018dota}, one of which is depicted in Fig. \ref{plane_sample_quantized}, along with its quantized versions with different bits\footnote{The IDs of image displayed in the figure along with five others are P1056, P0028, P0034, P0051, P0084 and P0121. These images were chosen for their high quality and extensive coverage area.}. In the simulations, each data rate range corresponds to a single quantized image. Higher data rates lead to higher-quality images, but also more interference with other UAVs.

For the purpose of comparison, we employ two benchmarks. First, the Bit-Oriented (BO) method, indicated by a blue line, operates based on the SAMA-D3QL algorithm but lacks semantic awareness. This method aims to maximize the total transmission rate of the UAVs without considering the quality of the reconstructed image on the server. Additionally, it does not account for the shared coverage degree in its observations (i.e., the term \(\bar{z}_{i}^{t}\) in \eqref{state_space}). This method is a relaxed version of the approaches proposed by Fu et al. \cite{fu2023uav} and Li et al. \cite{li2024blocklength}, representing an upper bound for their results in the current analysis. Second, a centralized heuristic algorithm (HU) illustrated by a orange line, serves as another benchmark by assigning each channel to the user with the highest path gain in a greedy fashion. Notably, in all scenarios, the upper bound i.e., the highest possible reconstructed image quality on the server without any system interference, is equal to one.

Moreover, it is worth noting that we did not restrict the problem formulation to specific UAV trajectories; in our simulation, we assume the UAVs hover along circular paths at designated velocities and directions, resulting in circles of varying radii, as illustrated in Fig. \ref{segment_modeling}. Importantly, the algorithm operates independently of the UAVs' future locations, ensuring that it does not depend on trajectory management.

\begin{table}[t!]
\caption{System Model Parameters.}
\begin{center}
\begin{tabular}{|c|c|}
\hline
\textbf{Parameter} & \textbf{Value} \\
\hline
Network area & 2D: $100 m \times 100 m$ \\
Noise power ($\sigma^2$) & $10^{-9}$ \\
Power levels ($P_{Q}$) & $[0, 5, 10] \ W$ \\
Default number of UAVs, BSs & $8, 2$ \\
Default number of channels & $3$ \\
Default UAV velocities & $\mathcal{U}(10, 20) \ m/s$ \\
Observation side length ($d$) & $\mathcal{U}(20, 40) \ m$ \\ \hline
LSTM history size ($\mathcal{H}$) & $4$ experiences \\
Capacity of experience memory & $1000$ experiences \\
Batch size & $64$ \\
Discount factor ($\gamma$) & 0.8 \\
Learning rate & $0.001$ \\
Exploration parameters $\widetilde{\epsilon}$, $\epsilon'$ & 0.001, 0.9995 \\
Approximator model & \begin{tabular}{@{}c@{}c@{}} LSTM with 64 units \\ + fully-connected layers \\ with $128$ and $64$ units \end{tabular}  \\
\begin{tabular}{@{}c@{}} Target network update frequency \\ \end{tabular} & Every $20$ steps \\
\hline
\end{tabular}
\label{t_prm}
\end{center}
\vspace{-15pt}
\end{table}

\subsection{results}
\subsubsection{Number of Channels}
Interference is the primary source of performance degradation in wireless networks, as it can significantly hinder the ability of devices to communicate effectively. By varying the number of available channels (denoted as \(\mathcal{C}\)), we can assess how the bounds of interference affect system performance. As illustrated in Fig. \ref{aggregated}-(A), a reduced number of channels leads to increased interference, which inhibits the UAVs' capacity to transmit high-quality images to the server. However, our approach demonstrated a considerably diminished negative impact in comparison to other methods. This resilience is attributed to the task-sharing strategy employed among UAVs that have overlapping observation areas. When UAVs collaborate effectively, sharing the burden of data transmission, they can mitigate the effects of interference.

\subsubsection{Number of UAVs}
As noted earlier, more UAVs can negatively impact performance evaluations due to heightened interference and the complexity of managing a larger fleet. Nevertheless, as shown in Fig. \ref{aggregated}-(B), this increase has a smaller adverse effect on our approach, akin to findings from the previous experiment. This robustness stems from the task-aware mechanisms within our framework, allowing UAVs to share responsibilities and optimize transmissions amid competing signals. Such characteristics highlight the scalable nature of our method, demonstrating its adaptability and capacity to maintain performance in dynamic environments with varying UAV numbers.

\subsubsection{UAV velocities}
\begin{figure}[!t]
\centerline{\includegraphics[width=2.8in]{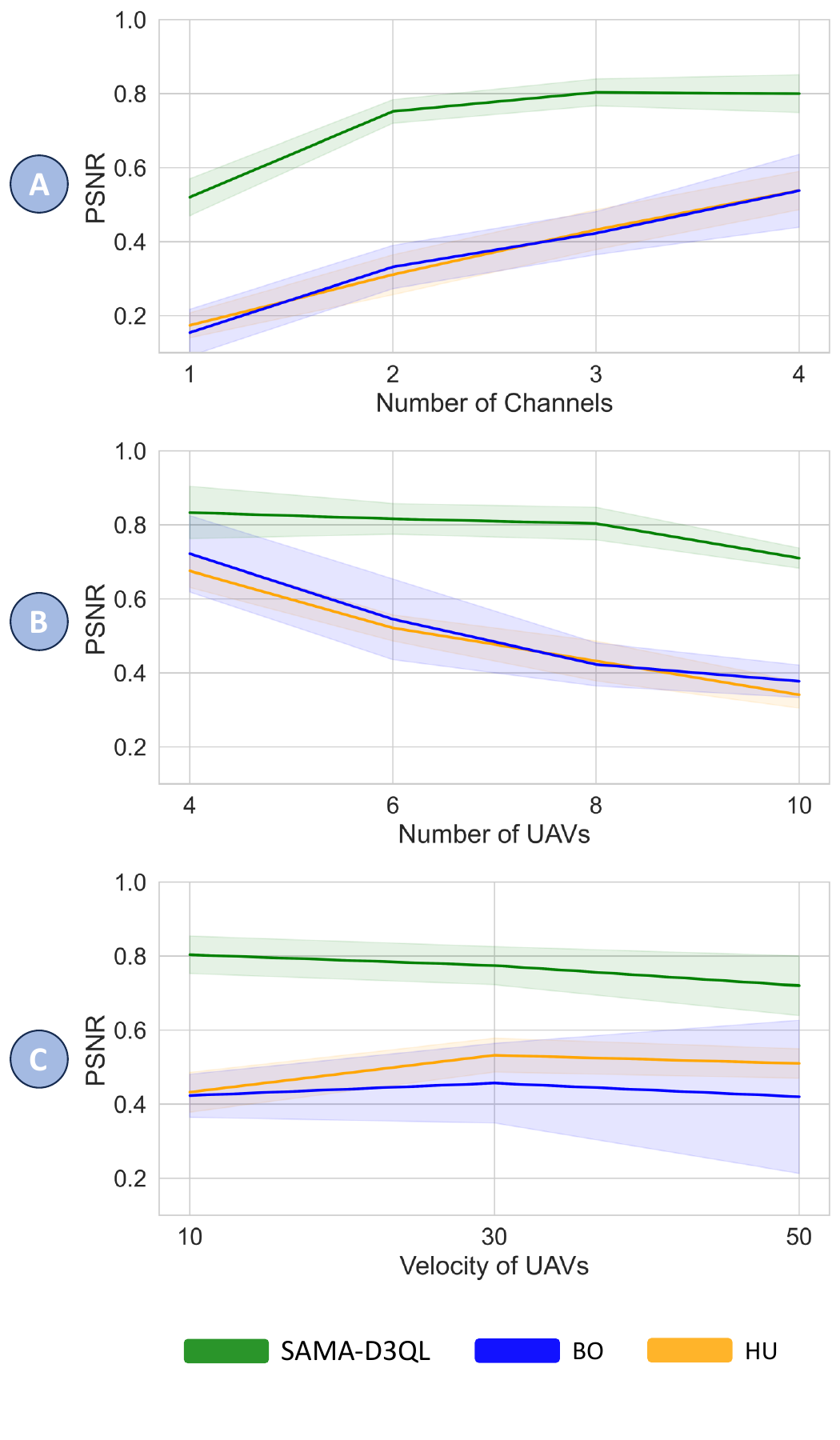}}
    \caption{ Average PSNR values for observation scenes of UAVs per (A) Number of channels (B) Number of UAVs (C) UAV velocities. Each line is an average of results over 6 plane images. Shaded areas represent the standard deviation of the values.}
    \label{aggregated}
\end{figure}

In this experiment, we modify the velocities of UAVs, directing them to complete a full circle during each episode. This adjustment indicates that increasing the velocities results in larger trajectory circles. Consequently, increased UAV velocities can be viewed as reflecting a higher level of dynamicity within the system, leading to states that are less similar across episodes and adding complexity to the learning process. Nevertheless, our approach continues to demonstrate high performance, indicating its adaptability to these changing dynamics. This resilience is illustrated in Fig. \ref{aggregated}-(C).

\subsubsection{Choice of Evaluation Metrics}
Throughout this paper, we utilized the PSNR metric to evaluate our scheme's performance, demonstrating how task-oriented communication policies are influenced by the selected evaluation metric. In this experiment, we trained the algorithm using the default parameters in Table \ref{t_prm}, first with PSNR-based rewards and then with Structural Similarity Index Measure (SSIM)-based rewards. Evaluation of the trained models with PSNR metric revealed that the first model achieved an average output of 0.80, while the SSIM-based model only reached 0.62. This discrepancy suggests that the SSIM-trained model did not effectively learn the necessary features for optimal performance, highlighting the importance of careful metric selection in our analysis.

\section{Conclusion}\label{s_con}
This paper leveraged the synergy among transformative technologies of 6G, to propose a dynamic and distributed power allocation scheme for semantic-aware networks, specifically focusing on multi-UAV area coverage as a use case. Our formulation and algorithm design accounted for the quality of observations while considering overlapping user observation areas. In this context, users are interconnected not only through interference but also through their source data. In future work, we will explore the problem further, including the prediction of image quality based on user data rates, which will facilitate the development of more efficient and sustainable power allocation algorithms. Additionally, we will draw upon our findings in \cite{shokrnezhad2024fairness} related to fairness and explainability. Implementing more advanced multi-agent learning algorithms that integrate global system information during training will also enhance system performance. Finally, predicting and orchestrating the required computational resources for 6G-based services \cite{farhoudi2023qos, 10816182, farhoudi2024discovery}, particularly Deep Neural Network (DNN)-based semantic encoding, enables a more comprehensive system view.

\section*{Acknowledgment}
The research work presented in this article was conducted in part at ICTFICIAL Oy. This work is partially supported by the European Union’s HE research and innovation program HORIZON-JUSNS-2023 under the 6G-Path project (Grant No. 101139172), and the European Union’s Horizon 2020 Research and Innovation Program through the aerOS project (Grant No. 101069732). The paper reflects only the authors’ views, and the European Commission bears no responsibility for any utilization of the information contained herein.

\bibliographystyle{IEEEtran}
\bibliography{IEEEabrv,main}

\end{document}